\documentstyle[aps,twocolumn]{revtex}

\begin{document}

\noindent  {\bf Reply: Van den Brink, Khaliullin and Khomskii}\\

In the preceding Comment~\cite{Shen}, Shen points out that the on-site Coulomb 
interaction, that can cause charge order in half-doped manganites~\cite{Brink99},
also destabilizes the magnetic CE-phase observed in these systems.

This is a valid observation, but it is not a priori clear whether in the relevant
parameter regime the C-phase is indeed lower in energy then the CE-phase within
our model~\cite{Brink99}. 
From the exact diagonalization of a 16-site ring, Shen finds a groundstate energy for 
the CE-phase as a function of $U$, which is compared to the energy of an {\it infinite} 
ring in the C-phase, in order to find a critical coupling $U_c=5t$ above which the
C-phase is stable. In order to determine the critical coupling, however, one should
handle finite size effects in both phases on equal footing and compare the CE-phase
energy with the energy of a 16-site C-phase ring, which is $E^C_{16}=-0.628t$. Using
this value for the energy, we can estimate 
from Fig.1 in Ref.~\cite{Shen} that $U_c \approx 10t$, which is in the relevant
parameter regime for the half-doped manganites.

As Shen states, other interactions which
are neglected in the model could very well compensate for the small energy 
differences between the C- and CE-phase. We wish to point out in particular
that coherent and incoherent hopping between two sites with opposite spin orientation 
lowers the kinetic 
energy of the CE-phase appreciably.

The main aim of our model is to explain the observed charge and orbital order,
{\it given} the CE magnetic structure. As to the stability
of the phases, indeed not all physical relevant effects are taken into account. Especially
the contribution of lattice distortions, which we neglected, are
important~\cite{Yunoki00}. Also, for instance, anharmonic effects that favor occupation 
of elongated orbitals~\cite{Khomskii00} that may stabilize the CE-phase, are not incorporated
in the model.

Moreover, one should also take into account that the hopping between two 
sites with antiferromagnetically coupled core spins is not completely blocked. 
For a finite Hund's rule coupling ($J_H$) an electron can hop coherently from a 
high spin state to lower spin states on a neighboring site, but 
even for a infinitely large $J_H$, there 
exist quantum effects which allow hopping that preserve the 
total high spin state $S+1/2$,
but with different projections $S^z=S+1/2$ and $S-1/2$ at different sublattices~\cite{Nagaev96}.
%These processes become more efficient for finite $J_H$. 
Such electron hopping
processes between C-type chains or between CE zigzag chains modify the total
energy.
%First, because 
%the Hund's rule coupling ($J_H$) is finite, and second because a hole can hop between to AF sites 
%when at the same time a magnon is emitted. 
One can easily estimate the coherent contribution to the
kinetic energy in second order perturbation theory. The contribution from hopping 
between two chains in the C-phase is then $\delta E_{C} = -\frac{3}{8} \frac{t^2}{\Delta}$
and in the CE-phase $\delta E_{CE} = -\frac{5}{8} \frac{t^2}{\Delta}$. This difference is
due to the fact that in the CE-phase also the $x^2-y^2$ orbital on the corner site
is partially occupied.
As typically
$\Delta \approx J_{H}$ and $J_H$ is in the order of $5t$, the energy lowering of
the CE-phase due to this effect would be larger than the energy loss due to the on-site 
Coulomb interactions in the CE-phase. The incoherent contributions that change the 
spin projections on different sublattices are more involved to calculate, but they would
certainly lower the energy of the CE-phase with respect to the C-phase as
the inter-chain hopping integrals are larger in the former.

%We therefore conclude that for a precise determination of the theoretical phase diagram 
%of the half-doped manganites it is necessary to incorporate many details which go beyond 
%the model of Refs.~\cite{Shen,Brink99}.\\

We therefore conclude that the model of Ref.~\cite{Brink99}, which correctly captures the interplay
of spin, charge and orbital degrees of freedom in the half-doped manganites and gives a 
reasonable description of their electronic structure, is by itself not sufficient for the precise
determination of the regions of stabilities of different phases. For this several other
factors, in particular those mentioned above, should be taken into account.\\

\noindent Jeroen van den Brink\\
{\small 
\indent Faculty of Applied Physics and MESA$^+$ Institute\\
\indent University of Twente\\ 
\indent Box 217, 7500 AE Enschede, The Netherlands}
\\

\noindent Giniyat Khaliullin\\
{\small 
\indent Max-Plack-Institute f\"ur Festk\"orperforschung\\
\indent Heisenbergstrasse 1, 70569, Stuttgart, Germany}
\\

\noindent Daniel Khomskii\\
{\small 
\indent Laboratory of Solid State Physics\\
\indent University of Groningen\\ 
\indent Nijenborgh 4, 9747 AG Groningen, The Netherlands}

\end{document}